\documentclass[a4paper,11pt]{article}
\pdfoutput=1 

\usepackage{jinstpub} 
\usepackage{epsfig}
\usepackage{float}

\title{\boldmath Performance of a First Microhexcavity Plasma Panel Detector with Muons}

\author[a,1]{A. Mulski,\note{Corresponding author.}}
\author[b]{Y. Benhammou,}
\author[b]{E. Etzion,}
\author[a]{C. Ferretti,}
\author[c]{P. S. Friedman}
\author[a]{D. S. Levin}
\author[b]{M. Raviv Moshe,}
\author[b]{D. Reikher,}
\author[a]{N. Ristow}

\affiliation[a]{University of Michigan, Department of Physics\\450 Church Street, Ann Arbor, Michigan 48109 USA}
\affiliation[b]{Tel Aviv University, Beverly and Raymond Sackler School of Physics and Astronomy\\Tel Aviv 69978, Israel}
\affiliation[c]{Integrated Sensors, LLC\\ Ottawa Hills, Ohio, 43606 United States}

\emailAdd{amulski@umich.edu}

\abstract{The microhexcavity plasma panel detector is a type of gaseous particle detector consisting of a close-packed array of millimeter-size hexagonal cells. The cells are biased to operate in Geiger mode where each cell functions as an independent detection unit. The response of the detector to ionizing radiation was investigated using low-energy radioactive $ \beta $ sources and cosmic ray muons. Efficiency measurements were conducted with cosmic ray muons in conjunction with a scintillator hodoscope. The rate response and signals obtained from the microhexcavity detector filled with  Penning gas mixture at atmospheric pressure are herein described. The relative pixel efficiency, after allowing for ion-pair formation in the gas volume, is 96.8 $ \pm $ 4.4$ \% $ for operation of the detector above an applied high voltage of 1000 V.}


\begin{document}
\maketitle
\flushbottom

\section{Introduction}
This article presents the first laboratory results from a microhexcavity plasma panel detector ($ \mu $Hex). The $ \mu $Hex consists of an array of close packed hexagonal pixels individually biased for a Geiger mode discharge. Each pixel functions as an electrically and optically isolated detection unit. Detectors with this \textit{closed}-cell architecture as well as \textit{open}-cell architecture devices have been under development for several years \cite{Ball:2014sia}, \cite{Ball:2014gqa}. The response of the $ \mu $Hex to low-energy $ \beta $ sources and efficiency measurements with cosmic ray muons are reported here. 

\section{Detector Description}
\begin{figure}[H]
\begin{center}
\epsfig{file=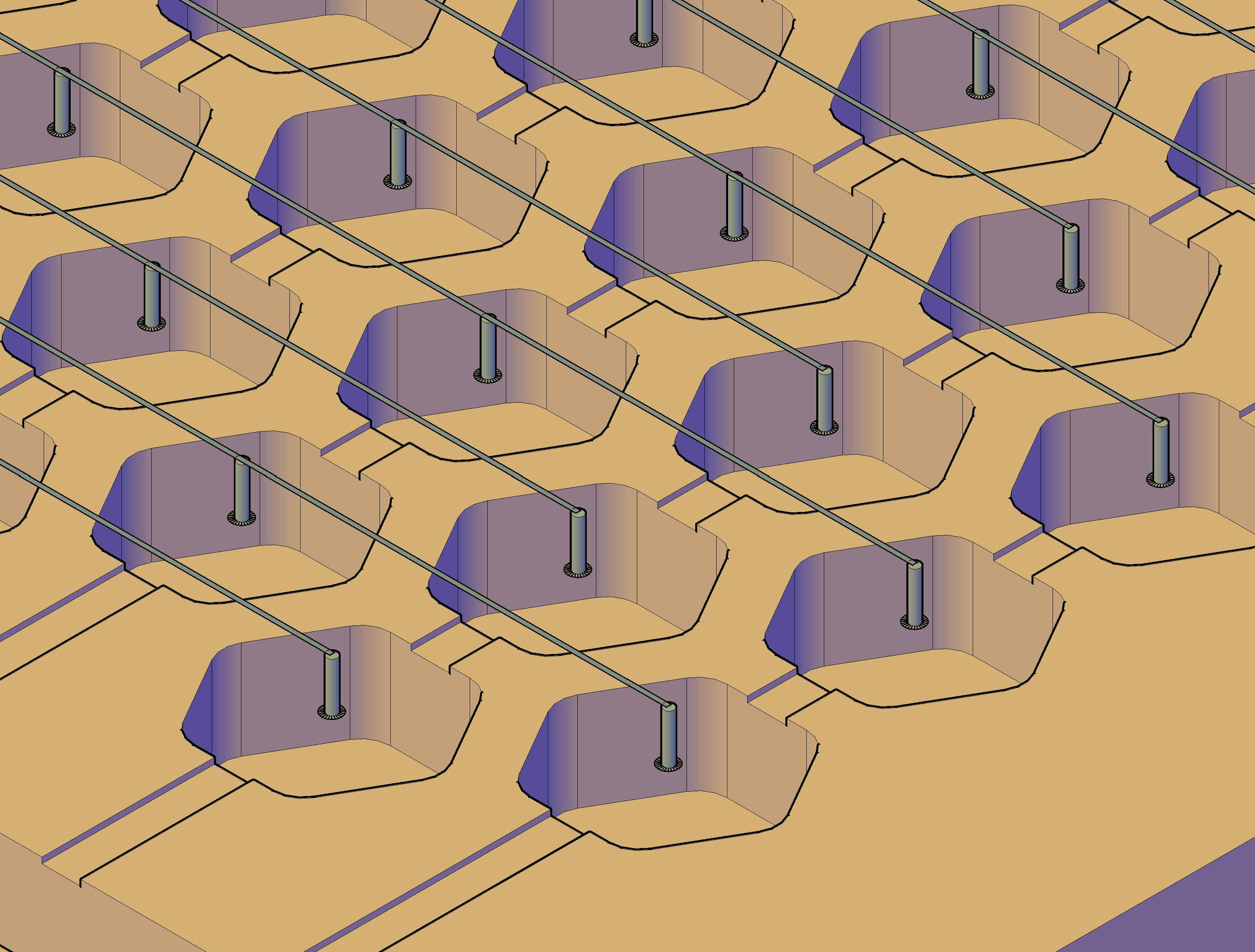,height=2 in}
\caption{Computer aided design rendering of $ \mu $Hex pixel array with a transparent cover plate. Gas channels are shown as shallow grooves that bridge from pixel to pixel. Circular RO anodes are connected to RO bus lines through conductive \textit{vias} that pass through the top substrate.}
\label{fig:pix}
\end{center}
\end{figure}

Pixels in the $ \mu $Hex consist of 2 mm wide regular hexagonal cavities fabricated 1 mm deep in a 1.4 mm layer of glass-ceramic dielectric substrate. The metallized inner surface of the cavity serves as the cathode of the pixel. This cathode is connected to a high voltage (HV) bus line through a conducting \textit{via} that runs from the cavity to the external bottom surface of the substrate. An external quench resistor isolates every cathode from the HV bus line. The top dielectric substrate (i.e. cover plate) hosts a circular readout (RO) anode above the center of each cavity, as shown in Figure \ref{fig:pix}. The anodes are connected through conducting \textit{vias} to thick-film printed RO bus lines on the top external surface of the detector. The two layers are then sealed together.

\vspace*{- 0.235 in}

\begin{figure}[H]
\begin{center}
\epsfig{file=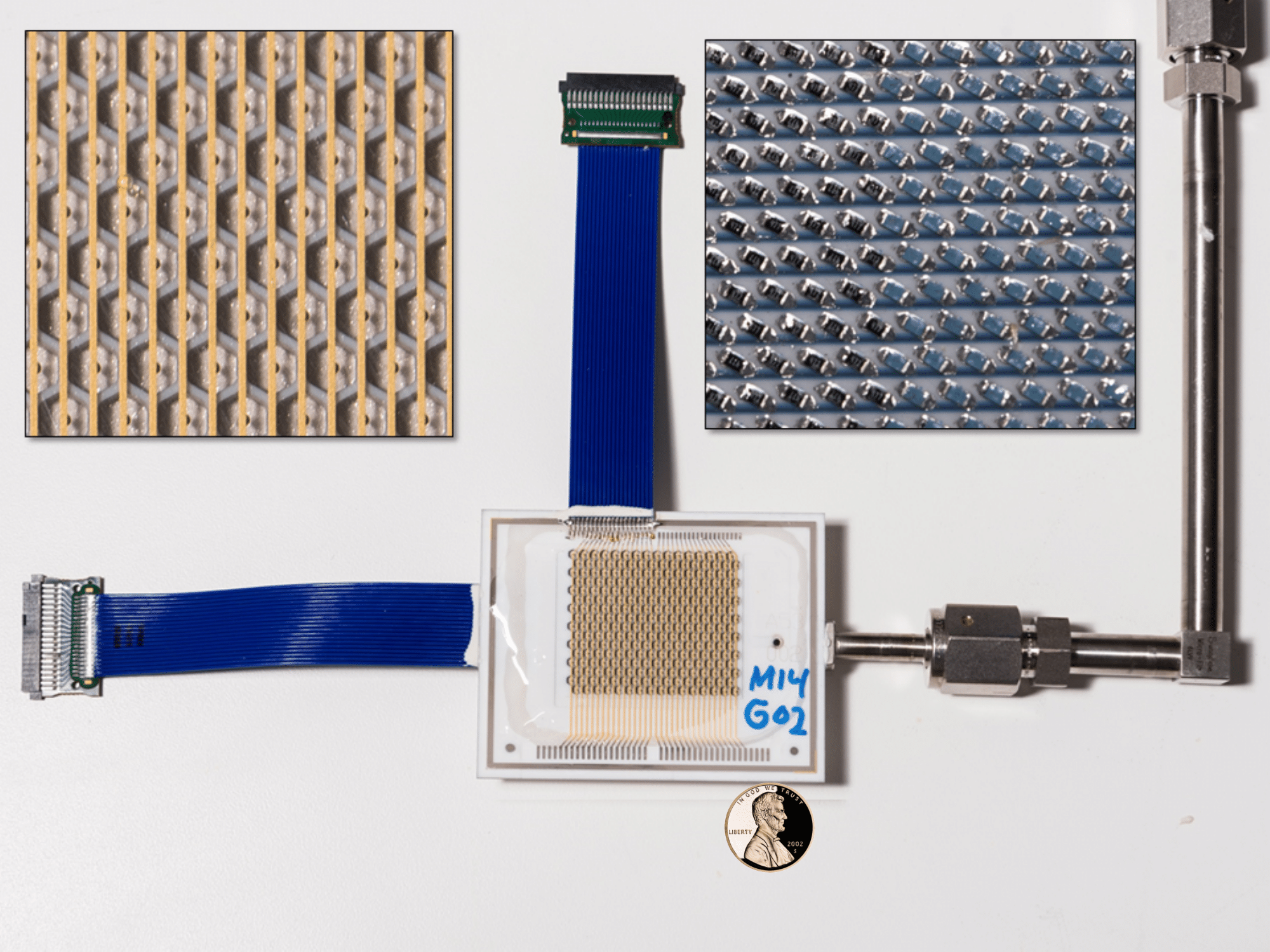,height=3.8 in}
\caption{A $ \mu $Hex detector with a transparent cover plate. The hexagonal pixel structure is visible below the cover plate, as well as the circular anodes with connecting readout lines (top left inset). The top right inset shows the external quench resistors situated at each pixel. The stainless steel gas port is used to evacuate and fill the device.}
\label{fig:uhexall}
\end{center}
\end{figure}

\setlength{\parindent}{0cm}Shallow channels machined in the lower substrate convey gas to each pixel. Neon-based Penning gas mixtures with an argon Penning component and an electronegative fluorocarbon component are used. A multi-channel gas mixing and pumping system is used to evacuate and fill the detector through a gas port on the side of the panel. Figure \ref{fig:uhexall} shows a $ \mu $Hex  with a glass cover plate and insets with detail of the cavity structure and external quench resistors.

\setlength{\parindent}{0cm}Each pixel is independently biased to undergo a Geiger mode gaseous discharge when an ion-pair is created in the cavity by a traversing charged particle. This event initiates a Townsend avalanche which occurs near the anode due to the non-uniform electric field. A simulation with a 3D electrostatic program (COMSOL Multiphysics) indicates the electric field is on the order of several MV/m. This high E-field leads to the formation of streamers and a gaseous discharge. As the charge difference between the electrodes begins to equalize, the E-field collapses and the discharge self-terminates. The signal produced is captured off the RO side of the panel from a 50 $ \Omega $ resistor between the anode bus lines and ground. 
\\~\\
Both spontaneous and secondary discharges are suppressed through the use of a Penning gas mixtures and external quench resistors. The Penning component of the gas fill de-excites long-lived metastable neon states. The electronegative component absorbs UV photons that prevents new photoelectric electrons from being emitted at the cathode surface.
\\~\\
The external quench resistor creates a recovery time which allows gas ions to neutralize before the pixel attains an E-field strength sufficient to support a new discharge. Recovery time is governed by the \textit{RC} time constant, where \textit{R} is the external quench resistance and \textit{C} is the effective pixel capacitance. The quench resistance is on the order of hundreds of M$ \Omega $s and the effective pixel capacitance measured between the cathode and anode using an LCR meter is less than 1 \textit{pF}. Single pixel recovery times are on the order of hundreds of microseconds.
\\~\\


\section{Experimental Results}
A $ \mu $Hex detector filled to 740 Torr with a Ne-Ar-CF$ _{4} $ gas mixture was used for the experiments described here. 16 RO lines with eight pixels per line were connected with parallel co-axial ribbon cables for a total of 128 instrumented pixels (125 were operational). The detector was staged on an optical test bench that allowed for careful alignment with a scintillator hodoscope. This detector staging was used for the efficiency measurements. To characterize the rate response, the rate was measured on 16 individual pixels. 

\subsection{Turn-On Voltage}
Beta particles from a $ ^{90} $Sr source were used to establish the voltage at which particles were first observed, known as the turn-on voltage. The HV was increased in 5 V increments until pulses were first observed on a digital sampling oscilloscope (DSO). The turn-on voltage was established to be ~900 V for this detector preparation. 

\subsection{Signal Characterization}
\begin{figure}[H]
\begin{center}
\epsfig{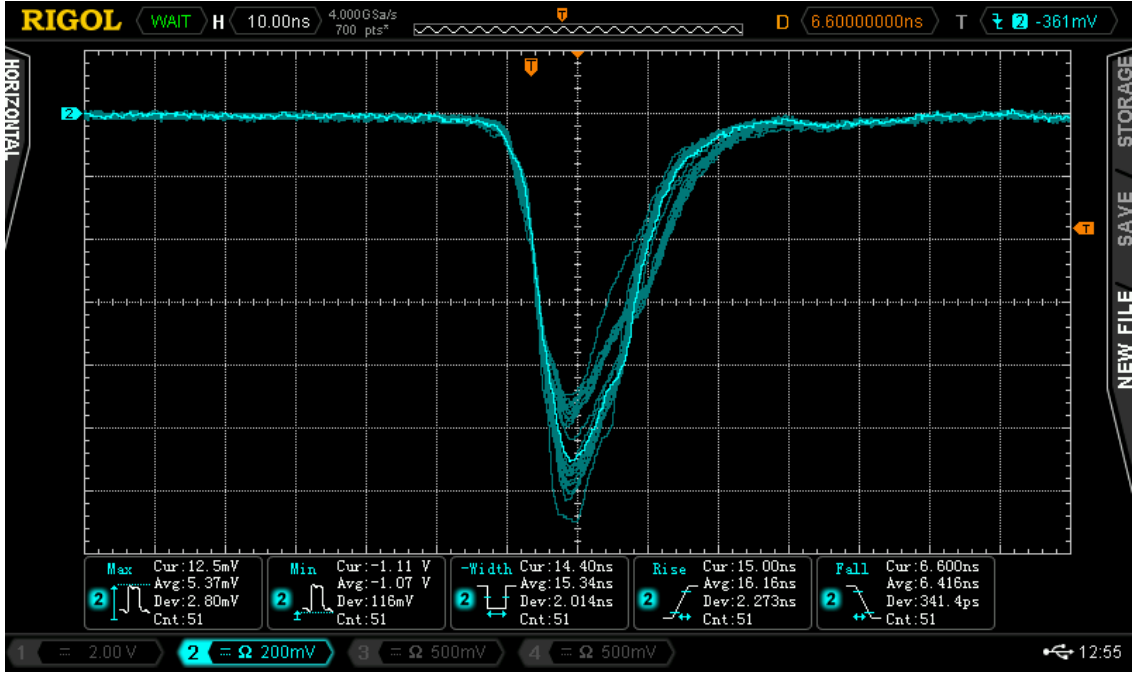}
\caption{An envelope of 51 signal pulses from the $ \mu $Hex detector in the presence of $ \beta $s from a $ ^{90} $Sr source. The x- and y-axis units are 10 ns per division and 200mV per division, respectively. The pulse shape is uniform across all pixels on the detector.}
\label{fig:pulse}
\end{center}
\end{figure}

The DSO image in Figure \ref{fig:pulse} shows an envelope of 51 pulses from a single RO line. Pulse characteristics are consistent from pixel to pixel with an average rise time of 7 ns and a FWHM of 15 ns. The pulse amplitude is an average of 1 V with a variation of $ \pm $ 30$\% $. Transients induced on neighboring RO lines are opposite in polarity with amplitudes on the order of 100 millivolts, roughly a factor of 10 less than the signal pulse.

\subsection{Response to Source}
\begin{figure}[H]
\begin{center}
\epsfig{file=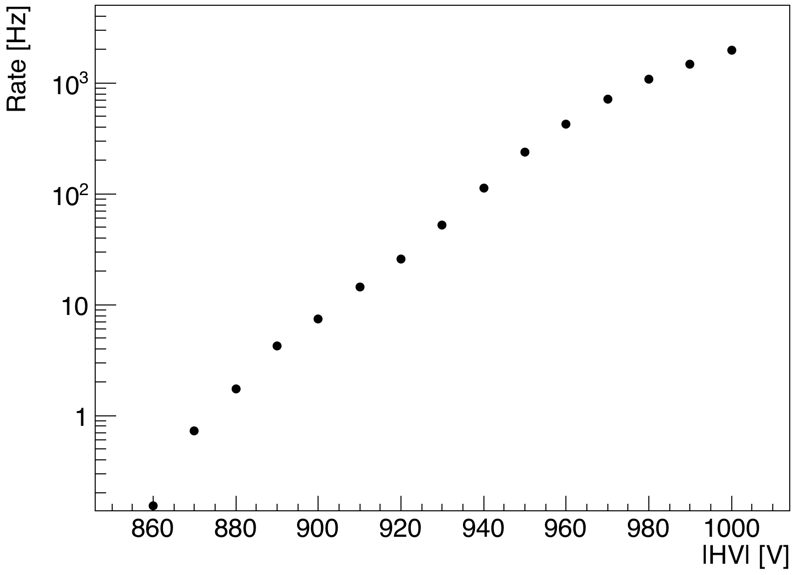,height=2.4 in}
\caption{$ \mu $Hex detector rate response as a function of HV with $ \beta $s from a $ ^{90} $Sr source. Each data point is the sum of the rate measured on 16 individual pixels.}
\label{fig:rate}
\end{center}
\end{figure}

Hit rate dependence on high voltage was measured with low-energy betas from an uncollimated $ ^{90} $Sr source. The source was positioned 20 cm above the $ \mu $Hex such that each RO line received a slightly different flux of $ \beta $ particles. The rate was measured at different values of applied high voltage beginning at the turn-on voltage and was increased in 10 V increments until 1000 V. 
\\~\\ 
Figure \ref{fig:rate} shows the $ \mu $Hex detector response as a function of HV. Each data point is the sum of the hit rates measured on 16 individual pixels over equal time intervals. The hit rate increases and begins to plateau as the voltage is increased. In the higher voltage range, residual rate increase is attributed to occasional after pulsing which occurs with a frequency of 10-30 percent of the total rate. In the absence of the source, the spontaneous discharge rate is less than 0.02 Hz per RO line at 1000 V. Assuming pixel response uniformity, the signal to spontaneous discharge ratio is $ \texttildelow $\textit{O(10$^{-4} $)}. 

\subsection{Efficiency with Cosmic Ray Muons}
\begin{figure}[H]
\begin{center}
\epsfig{file=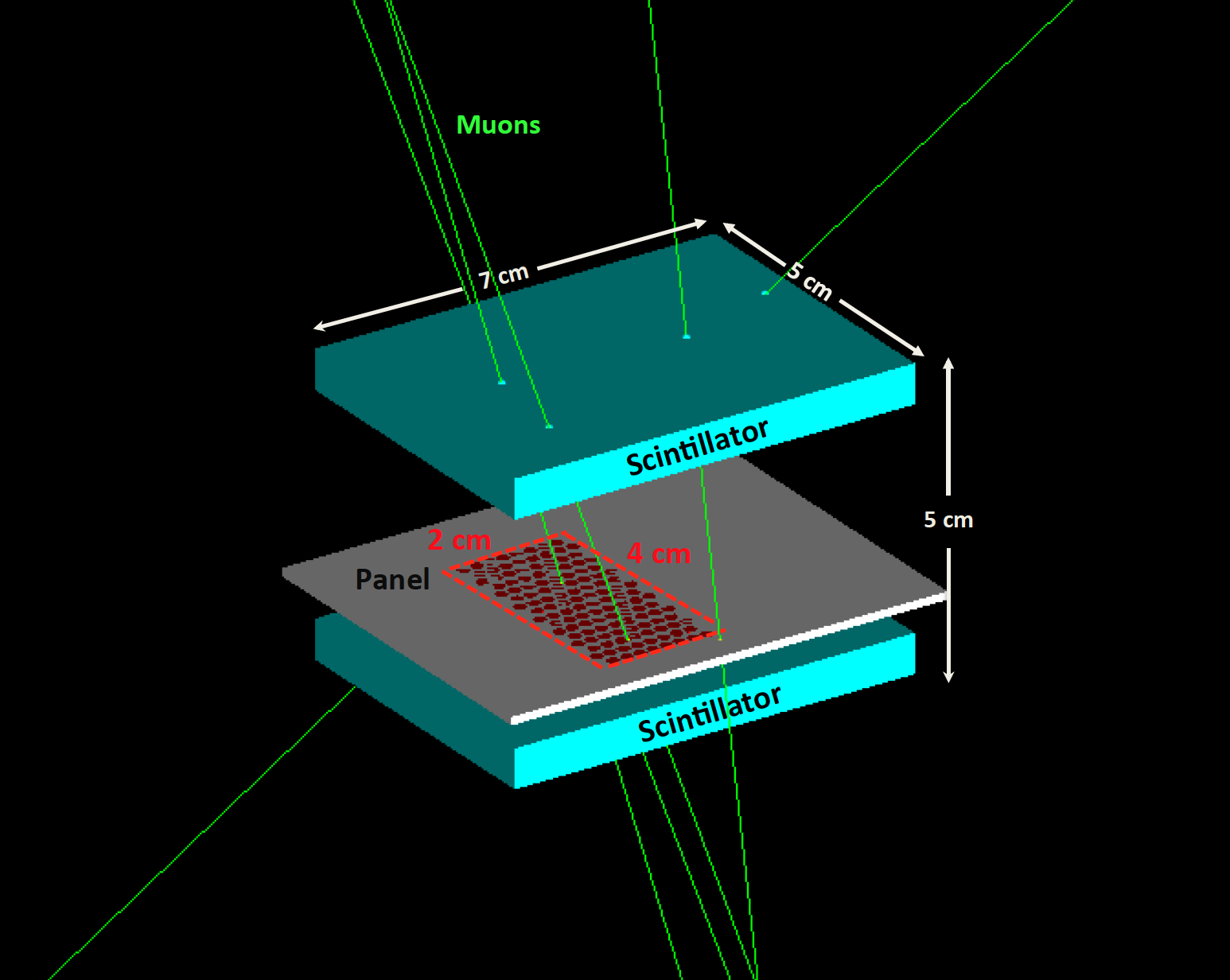,height=3 in}
\caption{Diagram of the experimental setup made in GEANT4. The active area of the $ \mu $Hex detector is outlined in red.}
\label{fig:setuplogic}
\end{center}
\end{figure}

A combination of GEANT4 Monte Carlo (MC) simulation and data was used to probe the efficiency of the $ \mu $Hex with minimum ionizing cosmic ray muons. In this study the pixel efficiency, denoted by $ \epsilon$, is the fraction of ionizing particle tracks passing anywhere in the pixel gas volume and generating at least one ion-pair that produce signals. The simulation models the setup of the experiment and was used to calculate the geometrical acceptance of the detector.
\\~\\
	For these efficiency measurements, the detector was mounted between a hodoscope consisting of two scintillator paddle detectors connected to photomultiplier tubes. An electronic level was used to establish parallelism between the detector and the scintillators within $ \pm $5 milliradians on the optical test bench. The scintillators were independently discriminated such that each scintillator gave comparable background rates at the same operating voltage.
\\~\\ 
	The timing coincidence of PMT hits within a 60 ns window was used to generate a hodoscope trigger. Signals from each RO line of the detector were discriminated before being sent to a multi-channel scaler. The threshold was set at 500 mV (roughly 50 $ \% $ of the average pulse amplitude) for all measurements. The geometrical setup for this measurement is shown in Figure \ref{fig:setuplogic}.

\setlength{\parindent}{0cm}The coincidence window between the $ \mu $Hex detector and the trigger was 1 microsecond. The measurement was not sensitive to secondary pulses on these timescales. Events were collected over a range of applied high voltages increased in 10 V increments. At each voltage, collection times ranged from a few hours to several days. 

\subsubsection{Monte Carlo Setup}
The geometric configuration was modeled in the MC toolkit GEANT4 version 10.02.p02 (2013) \cite{Geant4}. The $ \mu $Hex detector with its individual hexagonal pixels and the plastic scintillators forming the hodoscope trigger were recreated from both measurements and blueprint dimensions. Geantinos, a sterile particle provided by GEANT4, were used to simulate the acceptance of through-going muons. Particles were generated isotropically over the configuration with a cos$ ^{n} \theta$ angular distribution, where $ \theta $ is the polar angle and n is nominally taken to be 2 \cite{cosmic_dist_one}, \cite{cosmic_dist_two}, \cite{cosmic_dist_three}. The systematic effect of varying the parameter $ n $ is discussed in Section \ref{systematics}.
\\~\\
	The Poissonian probability distribution to produce an ion-pair is calculated based on ion-pair statistics for the specific gas fill and the incident particle's path length through the cavity gas volume. The formula used to calculate $ \epsilon $ is given by Equations \ref{eq:eff1} and \ref{eq:eff2}, where $ N_{H} $ is the number of hodoscope triggers (i.e. both scintillator paddles fire); $ N_{C} $ is the number of two-fold coincidences between the pixels and the hodoscope; $ A_{H} $ is the hodoscope angular acceptance; $ A_{C} $ is the combined acceptance of tracks passing through both a pixel and the hodoscope; $ \frac{dN}{d\Omega} $ is the differential rate of cosmic ray muons at sea level per cm$ ^2 $ per steradian taken to be cos$ ^{2} \theta$; $ \ell $ is the path length through the pixel gas volume, $ \mu $ is the gas dependent average number of ion-pairs per unit path-length, and $ P(n\geq 1, \mu \ell) $ is the Poissonian probability distribution for a through-going muon to produce a primary ion-pair in the cavity. The integrals in the numerator and denominator of Equation \ref{eq:eff2} are evaluated by the GEANT4 simulation. $\left ( \frac{N_{C}}{N_{H}} \right )_{Data} $ is measured directly, while $\left ( \frac{N_{C}}{N_{H}} \right )_{MC} $ is determined from the GEANT4 simulation.

\begin{equation}
\label{eq:eff1}
\epsilon = \frac{\left ( \frac{N_{C}}{N_{H}} \right )_{Data}}{\left ( \frac{N_{C}}{N_{H}} \right )_{MC} } 
\end{equation}	

\begin{equation}
\label{eq:eff2}
\epsilon \left ( \frac{N_{C}}{N_{H}} \right )_{MC} = \epsilon \frac{\int \frac{dN}{d\Omega}\cdot P(n\geq 1,\mu \cdot \ell)\cdot A_{C}(\theta,\phi)\cdot d\Omega}{\int \frac{dN}{d\Omega}\cdot A_{H}(\theta,\phi)\cdot d\Omega}
\end{equation}	

\setlength{\parindent}{0cm}$ \epsilon $ is proportional to the pixel acceptance relative to the hodoscope acceptance. The probability for a particle to produce a hodoscope trigger depends on the response uniformity of the two scintillator paddles to ionizing radiation. The spatial response uniformity was estimated with a radioactive $ \beta $ source and found to be within $ \pm $ 5$ \% $ over both paddles. A spatial response map was folded into the MC. The probability to produce a hit is also scaled linearly with track length up to the height of the scintillator. 
	
\subsubsection{Results}
\begin{figure}[H]
\begin{center}
\epsfig{file=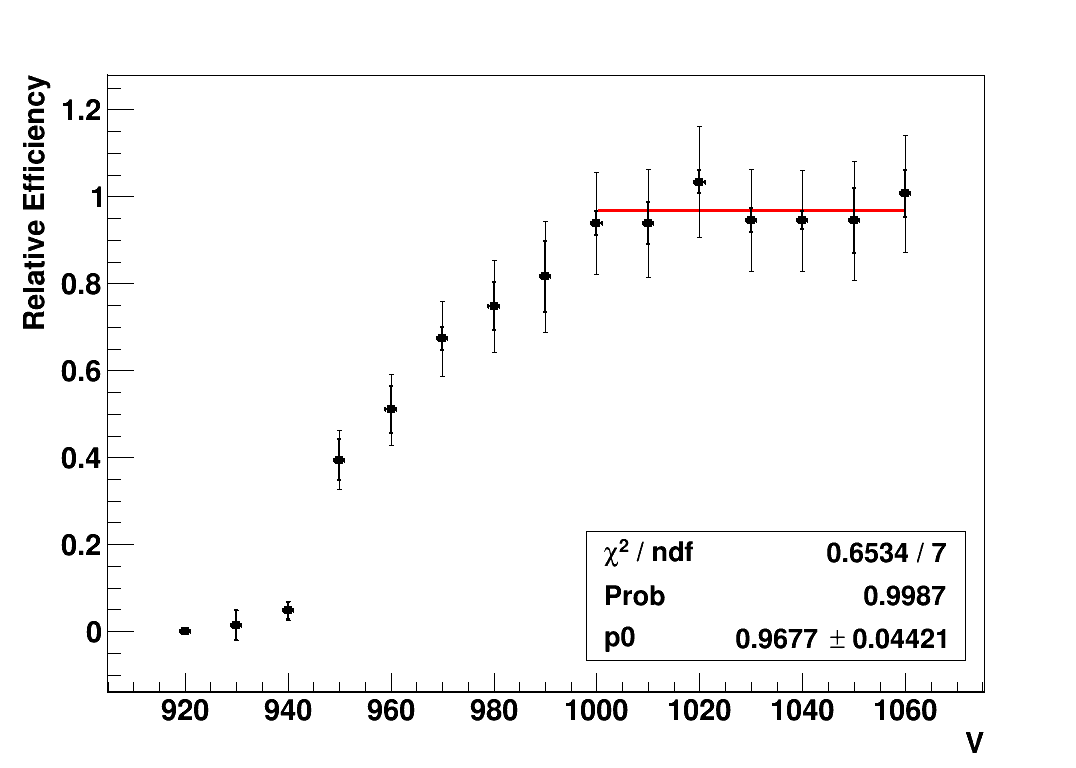,height=4 in}
\caption{The relative efficiency of the $ \mu $Hex detector as a function of voltage. The efficiency plateaus in a 60 V wide range of HV from 1000 V to 1060 V. A linear fit over this range was used to extract the relative efficiency $ \epsilon $.}
\label{fig:releff}
\end{center}
\end{figure}

About 15 percent of all measured hodoscope triggers had a coincidence with a pixel signal. The ratio of hodoscope triggers to two-fold coincidence between the hodoscope and the pixels obtained from the MC was $ \left( \frac{N_{C}}{N_{H}} \right )_{MC} $ = 14.7 $ \% $
\\~\\
The relative efficiency as a function of voltage is presented in Figure \ref{fig:releff}. The systematic error was calculated by the MC as discussed in section \ref{systematics}. The inner error bars in Figure \ref{fig:releff} are the statistical error on the measured ratio $ \left( \frac{N_{C}}{N_{H}} \right )_{Data} $ and the outer error bars are the addition in quadrature of the statistical and systematic errors.
\\~\\

\setlength{\parindent}{0cm}The response of the detector shows a plateau over a 60 V wide range from 1000 V to 1060 V. A fit over this range was used to obtain the relative efficiency $ \epsilon $ = 96.8 $ \pm $ 4.4$ \% $. The final uncertainty is a result of the fit error in which both systematic and statistical errors are included.

\subsubsection{Systematic Errors and Uncertanties} \label{systematics}
Four parameters were considered to contribute to the overall systematic error:

\begin{itemize}
  \item Detector placement
  \item Scintillator paddle response
  \item $ n $ in the cos$ ^{n} \theta$ cosmic ray angular distribution
  \item The number of primary ion-pairs per unit path length produced in the 	gas
\end{itemize} These errors were evaluated by independently varying each parameter in the MC while the others were held constant. The result of varying these parameters is summarized in Table \ref{tab:errors}. The systematics are reported as the percent difference in the value of $ \epsilon $ for each varied parameter.
\\~\\ 
	 The largest source of systematic error is associated with the scintillator paddle response as a function of path length. Accepting all particles incident on the scintillator paddles increased $ \epsilon $ by 1.37 $ \% $ and accepting none with a path length below the scintillator thickness (i.e. those generated with a large $ \theta $) decreased $ \epsilon $ by 8.65$ \% $.
\\~\\ 
	Changing the placement of the detector in the simulation varied $ \epsilon $ by 1.33$ \% $ when translated 2.5 mm horizontally and 0.66$ \% $ when translated 2.5 mm vertically. It was assumed the detector and scintillator paddles were configured such that any misalignment between them was negligible. The variation in position used to investigate this source of error was believed to be much greater than the actual error introduced from recreating the experimental setup.
\\~\\
	Different yields for the primary ion-pair per unit length produced in the gas are reported in the literature \cite{ionpair_one}, \cite{ionpair_two}. The MC was run with the two lowest and highest sets of values. The uncertainty introduced was established as half of the absolute difference between the two results. 
\\~\\
	The value of the n in the $ cos^{n}(\theta) $ term is nominally taken to be 2, but various sources report values of $ n $ ranging from 1.8 to 2.2 \cite{cosmic_dist_one}, \cite{cosmic_dist_two}, \cite{cosmic_dist_three}. To conservatively evaluate the error on the index $ n $, the simulation was run with a $ cos^{n}(\theta) $ angular distribution with $ n $ = 3, an increase from the nominal value by one integer. This variation resulted in differences of $ \epsilon $ on the order of a few percent. The systematic error is established symmetrically about the result as the uncertainty introduced from this variation.

\begin{table}[htbp]
\centering\

\smallskip
\caption{\label{tab:errors} Variations in the MC result for different sources of systematic error reported in $ \% $ differences of $ \epsilon $.} 
\begin{tabular}{| p{0.35\linewidth} | p{0.3\linewidth} | p{0.25\linewidth} |}
\hline
Varied Parameter& $ \epsilon $ Lower Systematic & $ \epsilon $ Upper Systematic\\
\hline
$ n $ in $ cos^{n}(\theta) $ & -2.35$ \% $ & 2.35$ \% $\\
\hline
Number of ion-pairs produced per unit path length in gas volume & -2.95$ \% $ & 3.15$ \% $ \\
\hline
Horizontal and vertical translation of detector & -1.35$ \% $ & 1.35$ \% $\\
\hline
Scintillator acceptence based on path length & -8.65$ \% $ & 1.37$ \% $\\
\hline
\end{tabular}

\end{table}

\section{Summary}
This article reports on the concept and first test results of the microhexcavity plasma panel detector. The pixels of the $ \mu $Hex detector function as isolated detection units. Each pixel is individually biased for a Geiger mode discharge when an ion-pair is created in the gas volume by a traversing charged particle. 
\\~\\
Initial tests with a low-energy beta source established the sensitivity of the $ \mu $Hex detector to ionizing radiation. The $ \mu $Hex detector did not generate signals in the absence of the source. Signal characteristics and rate response as a function of HV are uniform across all pixels. An efficiency study with MIPs assisted by a GEANT4 Monte Carlo simulation was used to determine the relative efficiency for a pixel to produce a signal when an ion-pair is created at any place in the gas volume by an incident charged particle. The relative pixel efficiency was found to be $ \epsilon $ = 96.8 $ \pm $ 4.4$ \% $ over a  60 V range of applied high voltage. The fabrication of a finer pitch, deeper microcavity plasma panel structure is currently underway. 

\section{Acknowledgments}
This research was supported by the United States National Science Foundation (NSF Grant  1506117), the United States-Israel Binational Science Foundation (BSF Grant No 2014716), the Abramson Center for Medical Physics, the I-CORE Program of the Planning and Budgeting Committee, and the Israel Science Foundation (Grant No.1937/12).

\end{document}